\documentclass[Journal]{IEEEtran}
\usepackage{amsfonts}
\usepackage{cite}
\usepackage{graphicx}
\usepackage{amsmath}
\usepackage{times}
\usepackage{latexsym}
\usepackage{mathrsfs}
\usepackage{dutchcal}
\usepackage{xcolor}
\usepackage{bm}
\usepackage{amssymb}
\usepackage{indentfirst}
\usepackage[skip=1pt,font=scriptsize]{caption}
\usepackage{titlesec}
\titlespacing*{\section}{0pt}{7pt}{3pt}
\titlespacing*{\subsection}{0pt}{3pt}{1pt}
\usepackage{stfloats}
\usepackage{cases}
\usepackage{array}
\usepackage{setspace}
\usepackage{fancyhdr}
\usepackage{color}
\usepackage{epstopdf}
\usepackage{balance}
\usepackage{amsthm}
\usepackage{diagbox}
\usepackage{mathrsfs}
\usepackage[ruled,linesnumbered]{algorithm2e}
\usepackage{subfigure}
\usepackage{etoolbox,amsthm}
\usepackage{wasysym}
\usepackage{bbm}
\usepackage{textcomp}
\usepackage{multirow}

\newcommand{\delete}[1]{}
\newcommand{\del}[1]{}

\makeatletter
\patchcmd{\thmhead@plain}%
  {\thmnote{ {\the\thm@notefont(#3)}}}% original form
  {\thmnote{ #3}}% new form
  {}{}
\let\thmhead\thmhead@plain
\patchcmd{\swappedhead}%
  {\thmnote{ {\the\thm@notefont(#3)}}}% original form
  {\thmnote{ #3}}% new form
  {}{}
\let\swappedhead@plain=\swappedhead
\makeatother
\IEEEoverridecommandlockouts
\begin{document}
%\title{False Data Injection Attacks Against Electric Vehicle Charging Station Voltage Regulation Capacity Estimation in Distribution System}
\title{False Data Injection Attack on Electric Vehicle-Assisted Voltage Regulation}

\author{Yuan Liu, Omid Ardakanian, Ioanis Nikolaidis, Hao Liang% <-this % stops a space
\thanks{This research was supported by funding from the Canada First Research Excellence Fund as part of the University of Alberta’s Future Energy Systems research initiative.} \and   %CFREF FES Grant Identification Number: CFREF-2015-00001.
\thanks{Y. Liu (email: yuan17@ualberta.ca) and H. Liang (email: hao2@ualberta.ca) 
are with the Department of Electrical and Computer Engineering, University of Alberta, Canada. 
O. Ardakanian (email: oardakan@ualberta.ca) and I. Nikolaidis (email: nikolaidis@ualberta.ca) are with the Department of Computing  Science, University of Alberta, Canada.}\vspace{-0.85cm}
}

\maketitle

\begin{abstract}
With the large scale penetration of electric vehicles (EVs) and the advent of bidirectional chargers, 
EV aggregators will become a major player in the voltage regulation market. 
This paper proposes a novel false data injection attack (FDIA) 
against the voltage regulation capacity estimation of EV charging stations,
the process that underpins voltage regulation in distribution system. 
The proposed FDIA takes into account the uncertainty in EV mobility and network conditions.
The attack vector with the largest expected adverse impact
is the solution of a stochastic optimization problem 
subject to a constraint that ensures it can bypass bad data detection.
We show that this attack vector can be determined 
by solving a sequence of convex quadratically constrained linear programs.
The case studies examined in a co-simulation platform, based on two standard test feeders,
reveal the vulnerability of the voltage regulation capacity estimation.
\end{abstract}


\IEEEpeerreviewmaketitle

\section{Introduction}
\IEEEPARstart{T}{he} notion of electric vehicle (EV) aggregators 
participating in the voltage regulation (VR) market 
has gained traction in recent years~\cite{vrbk}. 
Compared to traditional VR resources, 
such as thermal generators, on-load tap changers (OLTC), and capacitor banks, 
EV battery packs offer more advantages due to their inexpensive operation, 
high power capacity, and fast response~\cite{ebvr}.
The EV charging station (EVCS) agent, which is a typical aggregator, 
can coordinate charging and discharging of all EVs that are present in the station 
so as to respond to VR commands issued by the distribution system operator (DSO). 
Since the number of plug-in EVs and their state of charge (SOC) 
are highly variable and uncertain, 
the regulation capacity provided by an EVCS can vary over time. 
If this capacity is underestimated, the surplus will be wasted. 
% Conversely, if it is overestimated, 
On the other hand, compensating the shortage in VR capacity, due to mispredictions, 
either increases the operation cost or negatively affects the stability of 
the distribution network~\cite{mvpd}.
It is therefore vital for the DSO to accurately estimate the VR capacity
that can be provided by an EVCS in real time.
%to maximize the utilization of these economic VR resources. 

To facilitate the EVCS VR capacity estimation, 
the supervisory control and data acquisition (SCADA) system 
has been utilized in prior work for communications between the DSO and EVCS agents.
Reliable communication is indeed required to send the collected data for 
further analyses, such as state estimation, and bad data detection (BDD) and correction~\cite{rev3}. 
% With the deep integration of the physical and information layers, 
% For this reason, the vulnerability of the SCADA system to cyber attacks 
% has been the subject of many studies in recent years~\cite{dca}.
For this reason, cyber attacks targeting the communication system 
are regarded as one of the major threats to the reliable operation of power system~\cite{rev2}.
Examples of these attacks are false data injection attacks (FDIA), 
denial of service and replay attacks.
The FDIA could be more dangerous than other cyber attacks as it stealthier, 
enabling the attacker to disrupt the normal operation of the power system for a long time without being detected. 
Since first put forward in~\cite{fdiafirst}, 
such attacks have occurred several times, leading to long and catastrophic power outages. 
For example, in 2015, the attack launched against the Ukrainian power system's SCADA
caused a power outage affecting more than two million customers for six hours~\cite{ukr}. 
% Similarly, Israel's distribution system operator was targeted by a cyber attack in 2016~\cite{isr}, where the control center of DSO was hacked, 
% forcing the operators to shut down segments of the country's power grid.

Prior work has investigated FDIAs against distribution system state estimation (DSSE)
and proposed various methods to detect such attacks~\cite{stateest1,stateest2,stateest3,stateest4,stateest5,stateest6,ml}. 
%Load shifting attacks and the corresponding detection algorithm have been discussed in \cite{loadshift1}. 
% In \cite{smeter1} smart meter data manipulation 
% and algorithms for detecting it have been investigated. 
%Other types of attacks, such as topology attack~\cite{topology}, 
%network parameter alteration~\cite{netparaalt}, database %alteration~\cite{databasealt}, 
%and generation control attack~\cite{gencol} have been the subject of many studies too.
% In \cite{ml}, an extreme learning machine (ELM) framework 
% is proposed to detect FDIAs against state estimation. 
% The authors determine whether buses in a power system are under attack. 
However, none of these studies accounts for
the uncertainty of data communication. %in the SCADA system. 
Since not all false data injected by the attacker is received by the system operator 
in a timely fashion, existing FDIA strategies may not be as stealthy as proved theoretically.

Apart from misleading the DSSE process, the FDIA can affect ancillary services that are 
dependent on DSSE, e.g., VR, load shifting, and frequency regulation.
In \cite{dca}, the authors analyze how attackers can alter multiple field measurements 
in a coordinated manner to disturb voltage control algorithms. 
A machine learning-based two-stage approach is developed subsequently for detecting such attacks. 
Similarly, \cite{vcp} considers the impact of cyber attacks on VR 
in a distribution system with solar photovoltaics. 
An OLTC-induced FDIA against the VR process is investigated in \cite{stateest4}.
It is shown that it can lead to a wrong OLTC tap position,
causing serious under-voltage incidents.
Reference~\cite{essfdia} analyzes FDIA on battery energy storage 
installed in the distribution system, 
showing that the DSO can get completely wrong estimates of the battery energy content.
Reference~\cite{loadfdia} investigates the impact of FDIA on the load redistribution process. 
To identify the attacker in various scenarios, 
a novel cooperative vulnerability factor framework is introduced in~\cite{scad}, 
where each agent can track voltage fluctuations to perform accurate detection. 	
The impact of FDIA on remedial action schemes and the extent of failure propagation
have been studied in~\cite{fdiaimp}.
% moreover, the extent of failure propagation
% and cascading failures due to cyber attacks are investigated in that work. 
In~\cite{freqcontrol}, the authors propose a Luenberger observer 
and an artificial neural network-based approach 
to detect attacks against the load frequency control system.

Despite the vast literature in this area, 
there are still several challenges that are not fully addressed.
First, most related work relies on DC power flow, 
which is known to be inaccurate in certain distribution networks.
Second, an idealized communication model is usually adopted,
ignoring the impact of packet losses and delays on FDIA. 
Third, the related work does not investigate 
the vulnerability of the VR process to an FDIA
when the DSO does not receive the attack vector completely.
% especially the partly received attack vector should be analyzed. Hence, a novel FDIA construction method considering the stochastic communication process and the corresponding impact on the power system should be investigated based on DC power flow analysis.

This paper aims to address the gap that is identified above.
Specifically, we propose a novel approach for attack vector construction 
using a stochastic communication model. 
Our approach relies on AC power flow and a probabilistic model 
that captures the uncertainty of the EVCS up-regulation and down-regulation capacities.
We investigate the vulnerability of DSSE and BDD mechanism,
underpinning the VR process, to the proposed FDIA. 
The contribution of this paper is threefold:
\begin{itemize}
\item We develop a realistic VR model by extending DSSE and BDD 
with AC power flow analysis, and considering the locations of 
charging stations in the distribution network.
\item We propose an FDIA vector construction method
that relies on a stochastic communication model.
The optimal attack vector, which is the solution of a stochastic optimization problem,
can be found by comparing the solutions of a sequence of convex optimization problems.
% the corresponding construction algorithm is modified to improve the computation efficiency.
\item Through co-simulation, we expose the vulnerability of EVCS-assisted VR to the proposed FDIA. 
Our case studies, involving IEEE 33-bus and 123-bus test feeders,
suggest that the proposed FDIA is potentially more harmful 
than the traditional FDIA which relied on an idealized communication model.
This calls for the development of advanced BDD mechanisms that factor in varying network conditions,
a promising direction for future work.
\end{itemize}
The remainder of this paper is organized as follows. 
Section~II introduces the models, DSSE, and BDD process. 
Section~III presents the proposed attack vector construction method
under idealized and stochastic communication models.
Section~IV describes the case studies and the results obtained via co-simulation. 
Section~V concludes the paper and presents avenues for future work.

\begin{figure}
	\centering
	\includegraphics[width=\linewidth]{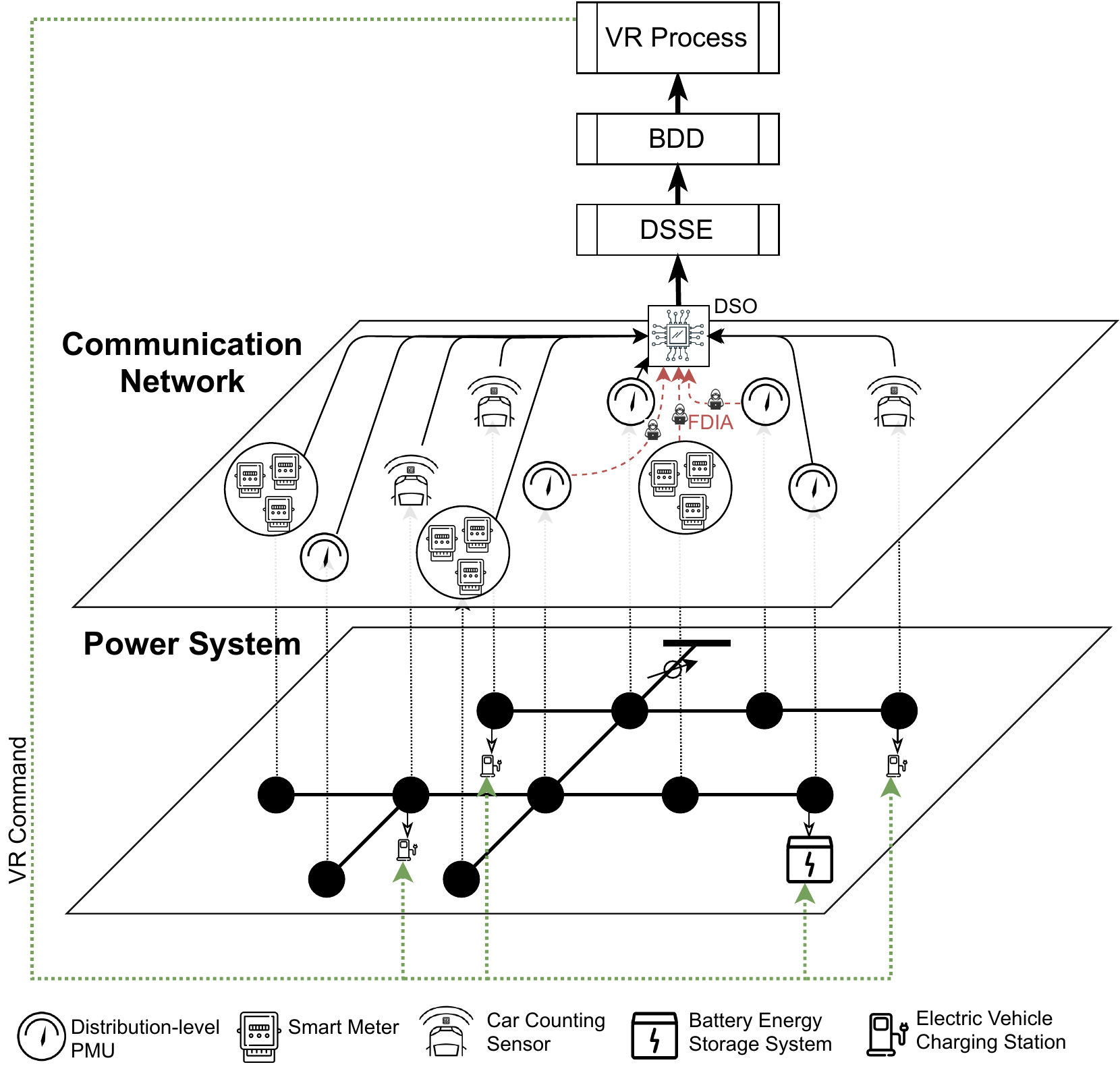}
	\caption{Illustration of FDIA against the VR process in a distribution system with EVCS and traditional VR resources, e.g., BESS. Red dashed lines show where the attacker injects false data.}
	\label{fig:systemmodel}
\end{figure}

\section{System Model}
Fig.~\ref{fig:systemmodel} shows a power distribution network
with various VR resources, such as battery energy storage systems (BESS), 
distributed generation units, and EV charging stations.
The distribution network is instrumented with a small number of distribution-level PMUs 
and numerous smart meters that are installed at customers' premises.
These sensors can be used to monitor the system state.
We assume sensor measurements are time-synchronized and sent at regular intervals 
(of length $\Delta t$) to the DSO control centre 
located at the substation through a communication network.
Similarly, the number of EVs parked at each EVCS is periodically sent to this control centre.

Due to random packet losses that could occur in the communication network 
and sparse deployment of distribution-level PMUs,
the DSO may have to use historical data to obtain pseudo measurements.
Together with real-time measurements, these pseudo measurements are used in the DSSE,
the output of which is sent through the BDD process to determine if it can be trusted.
If the estimated state passes the BDD process successfully, 
it will be utilized to issue control signals to the VR resources.  
Otherwise, the state estimate obtained by using pseudo measurements is utilized to issue control signals.

Given the reliance of DSSE on real-time measurements, %and pseudo measurements,
an attacker can target PMUs and EV counters for false data injection
to deceive the DSO into overestimating the VR capacity.
Yet, to minimize the risk of being detected, 
it should account for the stochasticity of data transmission 
that could result in pseudo measurements being used in lieu of real-time sensor measurements
when they are not received by the time DSSE is run.
We assume that the attacker has knowledge of the distribution system model, 
and unrestricted access to PMU measurements, EV counts, and 
the corresponding historical data that can be used to obtain pseudo measurements.
Furthermore, we make the following assumptions in this work:
\begin{enumerate}
	\item The DSO performs DSSE and sends VR commands at regular intervals;
	%the length of which is typically between 1 and 30 minutes~\cite{vrinterval,vrinterval2,vrinterval3};
	\item EVs take precedence over other VR resources in the VR process due to their responsiveness and low cost operation. Hence, they are always used first;
	\item Every EV that is plugged into a charger in an EVCS 
	is willing to act as a VR resource as long as 
	its charging demand is guaranteed to be satisfied before departure\footnote{A fraction of the revenue generated from the EVCS participation in VR can be distributed among 
	EVs to incentivize them to allow the EVCS to use their battery for voltage regulation. 
	That said, the design of this mechanism is outside the scope of this paper.};
% 	\item All the  distribution system  circuit information, PMU measurements, EVCS measurements, and corresponding historical data are known by the attacker;
% 	\item attacker can attack both PMU and EVCS measurements to let the DSO overestimate the VR capacity.
    \item There is enough parking stalls in each EVCS so that all EVs are admitted upon arrival. 
    But there might be a cap on the number of active chargers or the power that can be drawn simultaneously\footnote{The upper limit is imposed by the utility to avoid transformer overloading.}.
    The remaining parking stalls are occupied by EVs that are waiting for service;
	\item A car counter sensor is installed at each EVCS. 
	This sensor tracks the total number of EVs that are idle or being charged, without error.
\end{enumerate}
We now present the models adopted for voltage regulation, EVCS load, 
stochastic communication, BDD, and power flow.

\subsection{Voltage Regulation in Distribution System}
We consider a balanced three-phase distribution network 
with the set of nodes denoted $\mathbf N=\{1,\cdots,N\}$.
We denote the voltage magnitude, phase angle, active power injection, 
and reactive power injection of node $n$ at time $t$ by 
$\mathbf V_t=\{v_{n,t}\}$, $\bm \theta_t=\{\theta_{n,t}\}$, $\mathbf P_t=\{p_{n,t}\}$, $\mathbf Q_t=\{q_{n,t}\}$, respectively. 
Suppose there is a subset of nodes $\mathbf E = \{1,\cdots, E\}\subseteq \mathbf N$ 
where the connected load is an EVCS.
An EVCS at node $e\in \mathbf E$ can provide 
up-regulation capacity of $p^U_{e,t}$ and down-regulation capacity of $p^D_{e,t}$ at time $t$. 
We use the VR process described in~\cite{vrvol,vrev}. 
Specifically, the goal is to keep the average nodal voltage magnitude
within an acceptable range $[v^{\min},v^{\max}]$ around the reference voltage $v^R$
and preferably close to $v^R$.
Hence, the VR objective function is formulated as the squared deviation 
from the reference voltage averaged over all nodes:
\begin{subequations}\label{vobj}
\begin{align}
	&\min_{\{p'_{e,t}|e\in \mathbf E\}}\frac{1}{N}\sum_{n=2}^N(v^R-v_{n,t})^2\\
	&s.t.\ \ \ p_{e,t}-p'_{e,t}\leq p^D_{e,t}, e\in\mathbf E \\
	&\ \ \ \ \ \ \ p'_{e,t}-p_{e,t}\leq p^U_{e,t}, e\in\mathbf E\\
	&\ \ \ \ \ \ \ v^{\min}\leq v_{n,t}\leq v^{\max}, n\in\mathbf N,
\end{align}
\end{subequations}
where $p'_{e,t}$ is the active power contribution of EVCS $e$ when participating in VR.
Given the distribution network structure, 
it is possible that charging stations that offer the same up- and down-regulation capacities
but are connected to different nodes, contribute differently in the VR process.
Thus, we need to quantify the EVCS VR capacity 
%with a uniform standard, which can be calculated 
based on its impact on the voltage magnitude of node $n$ given by
\begin{align}\label{totalvr}
	\Delta v(n)=g^V_n(\mathbf P_t,\mathbf Q_t,\mathbf P^D_t,\mathbf P^U_t),
\end{align}
where $\mathbf P^D_t=\{p_{1,t}^D,\cdots,p_{E,t}^D\}$, $\mathbf P^U_t=\{p_{1,t}^U,\cdots,p_{E,t}^U\}$ 
are the sets of up- and down-regulation capacities of all EVCSs, 
$g^V(\cdot)$ is a function representing the maximum difference
in the voltage magnitude of node $n$ that could be
caused when the full VR capacity of the EVCS is utilized.
This function can be derived from power flow equations. 
The total VR capacity of the distribution network can be defined as
\begin{align}\label{vrcapacitity}
	\Delta v=\sum_{n\in \mathcal E}\Delta v(n),
\end{align}
where $\mathcal{E}\subseteq \mathbf N$ is a subset of nodes in the distribution network
that suffer from voltage limit violation problems. 
These nodes are typically at the end of distribution feeders.
% since the voltage magnitude decreases along the feeder if there is no energy injected.

\subsection{Stochastic Process for Characterizing the Number of EVs}\label{sec:queueing}
Suppose EVCS $e$ has $L_e$ charging points %$\mathbf L_e=\{1,\cdots,L_e\}$ 
and EVs arrive at this charging station following a Poisson process with rate $\lambda_e$.
Hence, the probability of having one arrival in a short time slot of length $\tau$
is $q_a=\lambda_e\tau+o(\tau)$ and the probability of no arrival in that time slot is $q'_a=1-\lambda_e\tau+o(\tau)$.
Suppose the EV charging times are independent and identically distributed 
exponential random variables with mean $1/\mu_e$.
Hence, the probability of having one departure from a specific active charger 
(i.e., charge service completion)
in a short time slot is $q_d=\mu_e\tau+o(\tau)$,
and the probability of no departure from that charger in that time slot is $q'_d=1-\mu_e\tau+o(\tau)$.
Note that $o(\tau)$ terms are negligible compared to $\tau$ 
when $\tau$ is sufficiently small; thus, they are ignored (infinitesimal asymptotics).
% We can model this EVCS as a $M/M/L_e$ queue. 

Let us denote the total number of charging and idling EVs in EVCS $e$ by $c_{e,t} \leq L_e$.
Hence, $\mathbf C_t=\{c_{1,t},\cdots,c_{e,t},\cdots,c_{E,t}\}$ is 
the set that contains the number of EVs that are in one of the $E$ stations at time $t$.
Assuming that there are $c_{e,t}=n$ EVs in EVCS $e$ at $t$, 
%the departure rate of that EVCS would be $c_{e,t}\mu_e$
we can derive the probability of having $n+i$ EVs in this EVCS at $t+1$ as follows:
\begin{align}
&P(c_{e,t+1}=n+i|c_{e,t}=n)=\\
\nonumber &\begin{cases}
q_a {q'_d}^n & i=1\\
q'_a q_d^n & i=-n\\
{n \choose i} q'_a q_d^{-i} {q'_d}^{n+i} + 
{n \choose i+1} q_a q_d^{1-i} {q'_d}^{n+i-1} & -n<i<1\\
0 & \text{otherwise}
\end{cases}
\end{align}
% \begin{align}
% &P(c_{e,t}=c_{e,t-1}+1|c_{e,t-1})=\lambda_e(1-\mu_e)^{c_{e,t-1}}\\
% &P(c_{e,t}=0|c_{e,t-1})=(1-\lambda_e)\mu_e^{c_{e,t-1}}\\
% \nonumber &P(0<c_{e,t}<c_{e,t-1}+1|c_{e,t-1})\\
% \nonumber &=\lambda_e\tbinom{c_{e,t-1}}{c_{e,t}-1}\mu_e^{c_{e,t-1}+1-c_{e,t}}(1-\mu_e)^{c_{e,t}-1}\\&\ \ \  +(1-\lambda_e)\tbinom{c_{e,t-1}}{c_{e,t}}\mu_e^{c_{e,t-1}-c_{e,t}}(1-\mu_e)^{c_{e,t}}
% \end{align}
where ${n \choose i}$ is a binomial coefficient. 
The DSO uses this probabilistic model to compute the most probable number of EVs in the EVCS, 
treated as the pseudo measurement,
when it does not receive the real-time measurement of the respective EV counter 
by the time it performs DSSE.

% In this work, to enable the DSO to estimate the real-time VR capacity of EVCSs, the real-time EV counts are collected by EV counting sensors and submitted to DSO. 

We now estimate the number of time slots in which an EV can participate in VR.
Suppose an EV arrives at the EVCS at $t$ with the initial SOC $s^I$, 
battery size $e^B$, charge (and discharge) power $p^C$, 
target SOC $s^T$ and expected parking time $t^P$. 
The expected charging time for this EV would be:
\begin{align}
	t^C=\begin{cases}
		\frac{(s^T-s^I)e^B}{p^C}, &s^I<s^T\\
		0,&s^I\geq s^T
	\end{cases}
\end{align} 
Naturally, EV owners seek to charge their car to the target SOC before departure. 
Once the current SOC reaches the target SOC, 
the EVCS aggregator will be able to charge or discharge the EV battery 
to provide up- or down-regulation. 
In this case, the corresponding charge/discharge power 
contributes to the VR capacity of the EVCS. 
However, to compensate for the energy withdrawn from the EV battery in the VR process, 
the battery must be recharged to the target SOC before the EV departure. 
Thus, the time slots available for providing VR capacity are as follows.
\begin{align}
	t^V=\begin{cases}
		\lfloor (t^P-t^C)/2\rfloor , &t^C<t^P \\
		0,&t^C\geq t^P
	\end{cases}
\end{align} 
where $\lfloor \cdot \rfloor$ is the floor function. 
Note that if the time required to satisfy the energy demand 
of an EV is longer than its parking time, 
the EV will be charged until departure without providing any VR capacity. 
Since the expected charging time and parking time are known in advance, 
the EVCS aggregator can flexibly control the EV charging process. 
In other words, the $t^V$ time slots can be utilized 
to provide the VR capacity when appropriate\footnote{It might be the case that 
some EV owners only allow their battery to be used 
for VR after their original energy demand is supplied. 
In that case, the time slots available to provide the VR capacity are fixed. 
Even in that case, the method proposed in this work is applicable as discussed in Section~\ref{sec:mdofified-dsse}.}. 
%However, considering various charging requirements, some strict EV owners prefer to finish the charging process before providing VR capacity so that they do not risk incomplete charging. Thus, their available VR time slots are fixed. In this work, we consider both flexible charging mode and strict charging mode to investigate the differences between them.

The EVs that arrive at the EVCS 
can have various initial SOCs, battery sizes, and parking times. 
We denote the probability density function for the initial SOC and parking time 
by $f_S(\cdot)$ and $f_P(\cdot)$ respectively.
Similarly, we denote the probability mass function for the battery size and 
maximum charge/discharge power
by $\varrho_B(\cdot)$ and $\varrho_C(\cdot)$, respectively.

\subsection{Stochastic Model for Data Transmission}
Suppose $U$ distribution-level PMUs are
installed at nodes $\mathbf U=\{1,\cdots,U\}$ 
to monitor the voltage magnitude and phase angle.
They send their measurements to the DSO at the end of each time slot,
which are then used to carry out DSSE 
$t$ time slots after the measurements are taken. 
The measurement packets that are not received by the control centre 
after $t$ time slots are deemed lost,
and the respective pseudo measurements will be utilized in the DSSE process.

Following the Gilbert–Elliott model,
we model changes in data transmission using a two-state Markov chain~\cite{GEmodel}.
In this chain, the good state (G) represents successful transmission,
and the bad state (B) represents unsuccessful transmission 
(i.e., the packet is not received after $t$ time slots).
The transition probability matrix of this chain is 
$$\begin{bmatrix}
	1-\kappa^{GB} & \kappa^{GB}\\
	\kappa^{BG} & 1-\kappa^{BG}
\end{bmatrix},$$
where $\kappa^{GB}$ is the probability of going from a good state to a bad state and $\kappa^{BG}$ is the probability of going from a bad state to a good state. 
Thus, the stationary distribution of this Markov chain can be written as
\begin{align}
	\pi(G)=\frac{\kappa^{BG}}{\kappa^{GD}+\kappa^{BG}}, \qquad
	\pi(B)=\frac{\kappa^{GB}}{\kappa^{GD}+\kappa^{BG}},
\end{align}
where $\pi(G)$ is the steady-state probability of being in a good state and 
$\pi(B)$ is the steady-state probability of being in a bad state.
This implies that in the steady state,
real-time PMU measurements are used in the DSSE process with probability $\pi(G)$ 
and pseudo measurements are used with probability $\pi(B) = 1 - \pi(G) $.

Based on the Gilbert-Elliott model, 
we can model the communication between sensors and the DSO control centre 
as independent Markov processes.
We use binary vectors $\bm \phi^U=[\phi^U_{1},\cdots,\phi^U_{U}]$ 
and $\bm \phi^E=[\phi^E_{1},\cdots,\phi^E_{E}]$
to collect the outcomes of data transmission 
in time slot $t$ for all PMUs and EV counters.
We use $\bm \phi=[\bm \phi^U,\bm \phi^E]^\top$ to compactly represent both vectors, the size of which is $U+E$.

\subsection{DSSE with BDD}
The DSSE concerns estimating the distribution system operating conditions 
given the measurement of a set of state variables~\cite{es},
e.g., nodal voltage magnitude and phase angle, 
real and reactive power injection at nodes, 
and real and reactive power flow in branches. 
The relation between the measurement $\bm{z}$ and the system state $\bm{x}$ is given by:
\begin{align}\label{obs}
	\bm{z}=h(\bm{x})+\bm{\varepsilon},
\end{align}
where $h(\cdot)$ is a nonlinear function that relates the measurement to the system state, 
and $\bm{\varepsilon}$ is the measurement noise with covariance matrix $\bm{R}$. 
We note that $h(\cdot)$ depends on the distribution system structure and line parameters.
% Generally, the objective of state estimation is to find an optimal system state $\bm{x}$ such that it can fit the measurement $\bm{z}$ best taking the impact of noise into account. In other words, 
Given $\bm{z}$, $h(\cdot)$, and $\bm{R}$, the system state $\bm{x}$ can be estimated 
by solving a weighted least squares (WLS) problem~\cite{wsl}:
\begin{align}
	\hat {\bm{x}}=\arg\min_{\bm{x}}[\bm{z}-h(\bm{x})]^\top\bm{W}[\bm{z}-h(\bm{x})],\label{orgprob}
\end{align}
where $W=\text{diag}\{\bm{R}^{-1}\}$ and $[\cdot]^\top$ denotes matrix transposition. 
This optimization problem can be solved via an iterative approximation method, 
such as the Newton-Raphson method.
However, nonlinearity of $h(\cdot)$ increases the computation overhead to a great extent. 
More importantly, its convergence cannot be guaranteed. 
Thus, this optimization problem is often simplified by linearizing the power flow equations. 
Consequently, $h(\cdot)$ is given by
\begin{align}
	h(\bm{x})=\bm{Hx},
\end{align}
where $\bm{H}$ is the measurement matrix obtained from the linearized power flow equations 
(we can also use the Jacobian matrix). 
In this case, the estimate can be derived as follows:
\begin{align}\label{estf}
	\nonumber\hat {\bm{x}}=&\arg\min_{\bm{x}}[\bm{z}-\bm{H}\bm{x}]^\top\bm{W}[\bm{z}-\bm{H}\bm{x}]\\
	=&(\bm{H}^\top\bm{WH})^{-1}\bm{H}^\top\bm{Wz}.
\end{align}
Considering the measurement noise of sensors and 
the possibility that measurements are modified by an attacker, 
DSOs typically employ a residual-based BDD mechanism to protect the DSSE process.
In this context, the residual can be defined as the difference 
between the actual measurement $\bm{z}$ and the measurement that corresponds 
to the estimated system state, i.e., $\hat{\bm{z}}=\bm{H}\hat{\bm{x}}$.
Then, by comparing the Euclidean norm of the residual $\bm{r}=\bm{z}-\hat{\bm{z}}$ 
against a threshold $\epsilon$, false data or erroneous measurement can be detected 
(if $\|\bm{r}\|_2>\epsilon$). 
Otherwise, the estimated system state $\hat{\bm{x}}$ can be trusted. 
The value of $\epsilon$ is typically determined by a hypothesis test $\text{P}(\|\bm{r}\|_2>\epsilon)<\tau$, where $\tau$ is the significance level.

In the VR process, the real-time system state is defined as $\bm x=[\mathbf P_t,\mathbf Q_t,\mathbf P^D_t,\mathbf P^U_t]^\top$ and the measurement vector is denoted by $\bm z=[\mathbf V_t,\bm \theta_t,\mathbf C_t]^\top$. 
The vectors $\mathbf V_t$, $\bm \theta_t$, and $\mathbf C_t$ collect
real-time measurements and pseudo measurements according to the communication result.

\subsection{Linear Power Flow Analysis}\label{sec:linpf}
We use a linear power flow analysis method to solve the DSSE problem more effectively. 
Let us denote the resistance and reactance between node $n$ and $k$ by $r_{nk}$ and ${x_{nk}}$, respectively. 
According to~\cite{LPFA}, we can approximate $\Delta\theta_{ik}\approx0$ and $v_i\approx1$~p.u. 
This allows us to linearize the power flow equations and write them in matrix form:
\begin{align}\label{lpf}
	\begin{bmatrix}
		\bm \theta'\\
		\bm v'
	\end{bmatrix}=
	\begin{bmatrix}
		\bm M^{2'} & \bm M^{1'}\\
		-\bm M^{1'} & \bm M^{2'}
	\end{bmatrix}^{-1}\!\!\!\!\Big(
	\begin{bmatrix}
		\bm p'\\
		\bm q'
	\end{bmatrix}-
	\begin{bmatrix}
		\bm m^{2c}\\-\bm m^{1c}
	\end{bmatrix}\theta_1-
	\begin{bmatrix}
		\bm m^{1c}\\\bm m^{2c}
	\end{bmatrix}v_1\Big),
\end{align}
where $\bm \theta'$, $\bm v'$ $\bm p'$ and $\bm q'$ are respectively the voltage phase angle, voltage magnitude, active power, and reactive power vectors of all the buses, except the slack bus. 
Here $\bm m^{1c}$ and $\bm m^{2c}$ are the first columns of $\bm M^1$ and $\bm M^2$ respectively, 
while
$\bm M^{1'}$ and $\bm M^{2'}$ are respectively the sub-matrices of $\bm M^1$ and $\bm M^2$ 
when we ignore the first column and the first row. 
Note that $\bm M^1$ and $\bm M^2$ are constant matrices obtained 
from the admittance matrix with their elements given by
\begin{subequations}
\begin{align}
	M^1_{nk}&=\frac{r_{nk}}{r^2_{nk}+x^2_{nk}},\qquad n\neq k,\\
	M^2_{nk}&=\frac{x_{nk}}{r^2_{nk}+x^2_{nk}},\qquad n\neq k,\\
	M^1_{nk}&=\sum_{k=\{1,\cdots,N\} - \{n\}}\frac{r_{nk}}{r^2_{nk}+x^2_{nk}},\\
	M^2_{nk}&=\sum_{k=\{1,\cdots,N\} - \{n\}}\frac{x_{nk}}{r^2_{nk}+x^2_{nk}}.
\end{align}
\end{subequations}

Based on \eqref{lpf}, the voltage magnitude and phase angle can be written as a linear function of $\bm P$ and $\bm Q$:
\begin{align}
	v_n=\sum_{k=2}^NM^2_{nk}p_k+\sum_{k=2}^NM^1_{nk}q_k+m^V_n,\label{lpfv}\\
	\theta_n=\sum_{k=2}^NM^2_{nk}p_k+\sum_{k=2}^NM^1_{nk}q_k+m^\theta_n,\label{lpftheta}
\end{align}
where $m^v_n$ and $m^\theta_n$ are a constant value given by
\begin{align}
	&[m^V_2,\cdots,m^V_N,m^\theta_2,\cdots,m^\theta_N]^\top\nonumber\\
	&=-
	\begin{bmatrix}
		\bm M^{2'} & \bm M^{1'}\\
		-\bm M^{1'} & \bm M^{2'}
	\end{bmatrix}^{-1}
	(
	\begin{bmatrix}
		\bm m^{2c}\\-\bm m^{1c}
	\end{bmatrix}\theta_1+
	\begin{bmatrix}
		\bm m^{1c}\\\bm m^{2c}
	\end{bmatrix}v_1
	).
\end{align}

\section{Stochastic FDIA on Voltage Regulation}
In the previous section, the system state has been augmented with 
the variables that are necessary for VR, 
namely up-regulation capacity, down-regulation capacity, and EV counts.
Given this new definition, a modified DSSE framework would be needed 
for the DSO to estimate the system states.
In this section, we first develop an optimization-based method 
for FDIA vector construction assuming an idealized communication model. 
We then present a new framework for DSSE that takes the VR variables into account.
This framework is utilized to formulate an optimization problem 
for FDIA vector construction under a stochastic communication model.

\subsection{FDIA against VR under Idealized Communication Model}
Under the idealized communication model, 
sensor data is guaranteed to be received before running DSSE, 
hence pseudo measurements are not used at all.
In this case, the attacker's objective to perturb measurements 
such that the adverse impact on the distribution network is maximized.
We define the impact of an FDIA by comparing the VR capacity before and after this attack:
\begin{align}
	\psi_{\bm\phi}(\bm \alpha)=\Delta v^A_{\bm\phi}-\Delta v_{\bm\phi},
\end{align}
where $\bm\phi$ is a 1-vector here because all packets must be received on time 
under the idealized model, 
$\Delta v_{\bm\phi}$ is defined in \eqref{vrcapacitity}, 
and $\Delta v^A_{\bm\phi}$ is the same quantity under FDIA. 
The best FDIA vector can be determined 
by solving the optimization problem below:
\begin{subequations}\label{subproblem}
\begin{align}
	&\max_{\bm \alpha}~\psi_{\bm\phi}(\bm \alpha)\label{detobj}\\
	&s.t.\ \ \bm z^A=\bm z+\bm \alpha\\
	&\ \ \ \ \ \ \hat {\bm x}^A=\Omega\bm z^A\\
	&\ \ \ \ \ \ \bm r^A=\bm z^A-\mathbf H\hat {\bm x}^A	\\
	&\ \ \ \ \ \ \|\bm r^{A}\|_2\leq \epsilon \label{ineqc}
\end{align}
\end{subequations}
Here $\bm \alpha^U=[\alpha^U_1,\cdots,\alpha^U_U]$ and $\bm \alpha^E=[\alpha^E_1,\cdots,\alpha^E_E]$ are the attack vectors concerning PMUs and 
EV counters (in charging stations); 
$\bm \alpha= [\bm \alpha^U,\bm\alpha^E]^\top$ is the combined attack vector;
$\bm z^A$ represents the modified measurements; $\Omega=(\mathbf H^\top\bm W\mathbf H)^{-1}\mathbf H^\top\bm W$ is the estimation matrix obtained from~\eqref{estf}. 
Note that the objective function is linear because 
we have linearized the power flow equations and estimation function. 

\subsection{Modified DSSE with VR Variables}\label{sec:mdofified-dsse}
Conventionally in DSSE, 
$h(\cdot)$ is obtained through power flow analysis
to calculate the voltage magnitude and phase angle measurements 
given the real and reactive loads.
% with the system state given by the active and reactive loads based on power flow analysis formulas. 
But when the VR variables are added to the state, 
a new function must be derived to estimate the VR capacity 
according to the extended system state, 
i.e., $\bm z=[\mathbf V_t,\bm \theta_t,\mathbf C_t]^\top$. 
% Notice that it is easier to estimate the EVCS VR capacity 
% according to the state of EVs than to estimate the state of EVs 
% according to the EVCS VR capacity. 
In this section, we first try to obtain the inverse of $h(\cdot)$ 
which relates the EV counts to VR capacity.
Knowing the inverse function, we derive $h(\cdot)$ at the end of this section.

Suppose one EV is parked at stall $\ell$ in EVCS $e$ in time slot $t$. 
This EV can be represented using a vector $[t_0,t^P,t^C,s^I,p^C,b^E]$, 
where $t_0\leq t$ is the arrival time slot and we have a constraint that $t-t_0\leq t^P$. 
Given $p_{e,t}$, which denotes the active load of EVCS $e$ obtained from the power flow analysis, 
we can estimate the expected number of charging EVs in this station at time $t$ as follows
\begin{align}
	c^C_{e,t}=p_{e,t}/\sum_{\forall p^C}p^C\cdot\varrho_C(p^C)
\end{align}
Accordingly, the expected number of idling EVs in this station 
can be calculated from $c^I_{e,t}=c_{e,t}-c^C_{e,t}$.

We consider two charging modes: \emph{strict} and \emph{flexible}.
The distinction between these two modes is whether or not
an EV can participate in VR before its initial charging demand is satisfied.
More specifically, 
a charging EV in the strict charging mode cannot provide up- or down-regulation capacity.
But, for an idling EV in the strict charging mode, 
the up- and down-regulation capacities are given by
\begin{align}\label{eq:down}
	p^{D}_{e,\ell,t}&=\begin{cases}
		p^C,& s^I+p^C\Delta t\leq s^T\\
		0, &\text{otherwise}
	\end{cases},\\\label{eq:up}
	p^{U}_{e,\ell,t}&=\begin{cases}
		p^C,&t^V>0 \\
		0, &\text{otherwise}
	\end{cases}. 
\end{align}
Turning our attention to the flexible charging mode,
the down-regulation capacity of a charging EV is zero. 
If a charging EV has abundant parking time, 
it can stop charging and immediately discharge its battery 
to provide the up-regulation capacity of $2~p^C$:
\begin{align}
	p^{U}_{e,\ell,t}&=\begin{cases}
		2~p^C, &t^P>t^C\\
		0, &\text{otherwise}
	\end{cases}.
\end{align}
% For an idling EV in the flexible mode, the up- and down-regulation capacities are given by
% \begin{align}
% 	p^{D}_{e,\ell,t}&=\begin{cases}
% 		p^C, & s^I+p^C\Delta t\leq s^T\\
% 		0,&\text{otherwise}
% 	\end{cases},\\
% 	p^{U}_{e,\ell,t}&=\begin{cases}
% 		p^C, &t^V>0\\
% 		0, &\text{otherwise}
% 	\end{cases}.
% \end{align}
The up- and down-regulation capacities of an idling EV 
are exactly the same as a charging EV (see Eq.~\eqref{eq:down}~and~\eqref{eq:up}). 

% where the EV can help with both up-regulation and down-regulation after finishing the charging process. 
% Besides, the up-regulation capacity can also be provided during the charging process. 
Given the probability density and mass functions, i.e.,
$f_S(\cdot)$, $\varrho_B(\cdot)$, $\varrho_C(\cdot)$, and $f_P(\cdot)$, 
our goal is to estimate the VR capacity of an arbitrary EV in an EVCS. 
To this end, we need to derive the probability distribution of $t^C$ 
based on the probability distribution of $b^E$, $p^C$, and $s^I$~\cite{transrand}:
\begin{align*}
\nonumber &f(t^C)\\&=\begin{cases}
    0, &t^C\notin (0,t^P]\\
    1-\int_0^{t^C}f_P(t^P)dt^P, &t^C=t^P\\
	\sum\limits_{b^E}\varrho_B(b^E)\sum\limits_{ p^C}\varrho_C(p^C)\int\limits_0^{s^T}f_S(s^I)\delta(\mathcal C^T)ds^I, &\text{otherwise}
\end{cases}
\end{align*}
where $\delta(\cdot)$ is the Dirac delta function and $\mathcal C^T=t^C-\frac{(s^T-s^I)b^E}{p^C}$.
%is the condition of the charging time.
The probability distribution of the flexible VR time can be calculated as follows:
\begin{align}
		\nonumber f(t^V)=\int_0^\infty f(t^P)\int_0^{t^P}f(t^C) \delta(\mathcal C^V)dt^Cdt^P,
\end{align}
where $\mathcal C^V=t^V-\lfloor (t^P-t^C)/2\rfloor$.
%is the condition of the available VR capacity.
Accordingly, the probability that an idling/charging EV has some available VR time slot is
\begin{align}
		\nonumber &P(t^V>0,t^C<t-t_0)\\
		&=\sum_{t^V}\int_0^\infty P(t^P)\int_0^{t-t_0}f(t^C) \delta (\mathcal C^V)dt^Cdt^P,\\
		\nonumber &P(t^V>0, t^P>t^C\geq t-t_0)\\
		&=\sum_{t^V}\int_0^\infty P(t^P)\int^{t^P}_{t-t_0}f(t^C) \delta (\mathcal C^V)dt^Cdt^P.
\end{align}
Given the expected EV charging time $t^C$ and arrival time $t_0$, 
the probability that it is idling at time $t$ is $P(t^C<t-t_0)$ which is given by
\begin{align}
	P(t^C<t-t_0)=\int_0^{t-t_0}f(t^C)dt^C,
\end{align}
and the probability that it is charging is $P(t^P>t^C\geq t-t_0)$ which is given by
\begin{align}
	P(t^P>t^C\geq  t-t_0)=\int_{t-t_0}^{t^P}f(t^C)dt^C.
\end{align}
Observe that $t^C< t-t_0$ and $ s^I+p^C\Delta t\leq s^T$ are not independent events. 
We derive the conditional probability $P(t^C|s^I,p^C)$ as follows:
\begin{align}
		P(t^C|s^I,p^C)=\begin{cases}
		\sum_{b^E}P^E(b^E)\bm I_{\mathcal C^T=0}(b^E), &0<t^C<t^P\\
		1-\int_0^{t^P}f(t^C)dt^C, &t^C=t^P\\
		0, &\text{otherwise}
	\end{cases}
\end{align}
where $\bm I_{\mathcal C^T=0}$ is the indicator function of set 
$\{b| \mathcal C^T=0\}$, i.e., it is equal to 1 when $b^E$ satisfies the equation
$\mathcal C^T=0$, and is 0 otherwise. We can write:
% This allows us to determine $P(t^C< t-t_0, s^I+p^C\Delta t\leq s^T)$:
\begin{align}
		&P(t^C< t-t_0\  ,  s^I+p^C\Delta t\leq s^T)\\
				\nonumber&=\int_0^{t-t_0}P(t^C|s^I,p^C)\sum_{\forall p^C}P(p^C)\int_0^{s^T}s^I\delta(\mathcal C^T)ds^Idt^C.
\end{align}
Based on the analysis above, the expected VR capacity of an EV in the strict charging mode 
is given by
\begin{align}
	\bar p^{ID}_{e,\ell,t}&= p^C\frac{P(t^C< t-t_0, s^I+p^C\Delta t\leq s^T)}{P(t^C<t-t_0)},\\
	\bar p^{CD}_{e,\ell,t}&= 0,\\
	\bar p^{IU}_{e,\ell,t}&= p^C\frac{P(t^V>0,t^C<t-t_0)}{P(t^C<t-t_0)},\\
	\bar p^{CU}_{e,\ell,t}&= 0,
\end{align}
and the expected VR capacity of an EV in the flexible charging mode is given by
\begin{align}
		\bar p^{ID}_{e,\ell,t}=~&p^C\frac{P(t^C< t-t_0, s^I+p^C\Delta t\leq s^T)}{P(t^C<t-t_0)},\\
	\bar p^{CD}_{e,\ell,t}=~&0,\\
	\bar p^{IU}_{e,\ell,t}=~&p^C\frac{P(t^V>0, t^C<t-t_0)}{P(t^C<t-t_0)},\\
			\nonumber\bar p^{CU}_{e,\ell,t}=~&p^C\frac{P(t^V>0,t^C<t-t_0)}{P(t^C<t-t_0)}\\&+2p^C\frac{P(t^V>0, t^P>t^C\geq t-t_0)}{P(t^C<t-t_0)}.
\end{align}
where $\bar p^{ID}_{e,\ell,t}$ and $\bar p^{IU}_{e,\ell,t}$ 
represent the down- and up-regulation capacities of an idling EV, 
and $\bar p^{CD}_{e,\ell,t}$ and $\bar p^{CU}_{e,\ell,t}$ 
represent the same quantities for a charging EV.
Putting it all together, the total VR capacity of an EVCS can be estimated 
using a linear function of the total number of charging and idling EVs in that station:
\begin{align}
	\bar p^D_{e,t}=\bar p^{ID}_{e,\ell,t}c^I_{(e,t)}+\bar p^{CD}_{e,\ell,t}c^{C}_{(e,t)}\label{dvr},\\
	\bar p^U_{e,t}=\bar p^{IU}_{e,\ell,t}c^I_{(e,t)}+\bar p^{CU}_{e,\ell,t}c^{C}_{(e,t)}.\label{uvr}
\end{align}
Recall that the voltage magnitude and phase angle of each node can also be calculated 
from \eqref{lpfv} and \eqref{lpftheta}, which are linear. 
Thus, the inverse of $h(\cdot)$ is a linear function of measurement $z$, 
and can be written in matrix form. This allows us to write $h(\cdot)$ in matrix form too.

\subsection{FDIA against VR under Stochastic Communication Model}
Due to stochastic packet drops in the communication network, 
the measurement sent by PMUs may not be received by the DSO when it attempts to run DSSE,
causing the respective pseudo measurements to be used in the process instead.
Let us denote the measurement vector utilized by the DSO in DSSE by 
$\bm z^R=\bm \phi \bm z+(\mathbf 1-\bm\phi)\bm z^P$, 
where $\textbf{1}$ is the 1-vector, 
$\bm\phi$ is a binary vector that indicates sensor measurements 
that are successfully received by the DSO,
and $\bm z^P$ is the pseudo measurement vector.
Similarly, we define
$\bm z^{AR}=\bm \phi \bm z^A+(\mathbf 1-\bm\phi)\bm z^P$ 
to be the measurement utilized in DSSE when FDIA is performed. 
We write the joint probability distribution of the communication results as
\begin{align}
	P(\bm \phi)=\prod_{\{i|\bm \phi_i=1\}}\pi(G)\prod_{\{j|\bm \phi_j=0\}}\pi(B),
\end{align}
and the probability of receiving measurement vector $\bm z^R$ as
\begin{align}
	P(\bm z^R)=\sum_{ \{\bm\phi|\bm z^R=\bm \phi \bm z+(\mathbf 1-\bm\phi)\bm z^P\}}P(\phi).
\end{align}
Hence, the probability of obtaining a state estimate $\hat{\bm x}$ from DSSE 
given the measurement $\bm z^R$ is
\begin{align}
	P(\hat{\bm x})=\sum_{\{\bm z^R|\Omega\bm z^R=\hat{\bm x}\}}P(\bm z^R),
\end{align}
and the corresponding residual is
\begin{align}
	\bm r=\bm z^R-\hat{\bm z}=\bm z^R-\mathbf H\Omega\bm z^R.
\end{align}
The probability distribution over the residuals can be calculated as follows
\begin{align}
	P(\bm r)=\sum_{\{\bm z^R|\bm z^R-\mathbf H\Omega(\bm z^R)=\bm r\}}P(\bm z^R).
\end{align}

Recall that the total VR capacity can be calculated in terms of voltage magnitude differences
according to %\eqref{totalvr} and 
\eqref{vrcapacitity}.
From \eqref{lpfv} and \eqref{lpftheta}, 
the voltage magnitude difference can be calculated 
as a linear function of node active and reactive loads 
along with up- and down-regulation capacities of every EVCS. 
The up- and down-regulation capacities themselves
are linear functions of EV counts as shown in~\eqref{dvr} and \eqref{uvr}. 
Thus, the $\Delta v$ term in \eqref{vrcapacitity} is a linear function 
of the DSSE result, $\hat{\bm x}$, and can be written as
\begin{align}
	\Delta v=\mathcal V\Omega \bm z,
\end{align}
where $\mathcal V$ can be derived from \eqref{vrcapacitity}, \eqref{lpfv}, \eqref{lpftheta}, \eqref{dvr}, and \eqref{uvr}.
This yields a probability distribution over the VR capacity
\begin{align}
	P(\Delta v)=\sum_{\{\hat{\bm x}|\hat{\bm x}=\Delta v\}}P(\hat{\bm x}).
\end{align}
Following the same approach, we can obtain the DSSE result given an attack vector $\bm \alpha$. 
We add $A$ to the subscript to mark the variables related to FDIA.

Given the derivations above, we formulate an optimization problem
for the FDIA vector construction under the stochastic communication model:
\begin{subequations}\label{original-problem}
\begin{align}
	&\max_{\bm \alpha}\Psi(\bm \alpha) \hfill &\label{stoobj}\\
% 	&\max_{\bm \alpha} P(\Delta v^{A}_{\phi_1})\psi_{\phi_1}(\bm \alpha)+P(\Delta v^{A}_{\phi_2})\psi_{\phi_2}(\bm \alpha) \label{stoobj}\\
	&s.t.\ \ \Psi(\bm \alpha)=\sum_{\{\bm \phi\}}\eta_{\bm\phi}(\bm\alpha)P(\Delta v^{A}_{\bm\phi})\psi_{\bm\phi}(\bm\alpha)&\\
	&\ \ \ \ \ \ \hat{\bm x}^{A}_{\bm\phi}=\Omega\bm z^{AR}_{\bm\phi}&\forall \bm\phi\\
	&\ \ \ \ \ \ \Delta v^A_{\bm\phi}=\mathcal V\hat{\bm x}^A_{\bm\phi}&\forall \bm\phi\\
	&\ \ \ \ \ \ \bm r^{A}_{\bm\phi}=\bm z^{AR}_{\bm\phi}-\mathbf H\hat{\bm x}^A_{\bm\phi}\label{residual}&\forall \bm\phi\\
	&\ \ \ \ \ \ \eta_{\bm\phi}=\begin{cases}
		1,  &\|\bm r^{A}_{\bm\phi}\|_2\leq \epsilon\\
		0, &\text{otherwise}
	\end{cases}&\forall \bm\phi\label{detect}
\end{align}
\end{subequations}
where $\eta_{\bm\phi}(\bm\alpha)$ is the BDD result 
associated with attack vector $\bm\alpha$ and communication result $\bm\phi$.
The optimal point of this problem is the attack vector that has 
the largest expected adverse impact on the VR process.
Notice that, for a specific communication result $\bm \phi$, $P(\Delta v^A_{\bm\phi})$ is constant 
and $\psi_{\bm\phi}(\bm\alpha)$ is a linear function of the attack vector $\bm\alpha$. 
Since $\eta_{\bm\phi}$ depends on both the attack vector $\bm \alpha$ 
and the communication result vector $\bm\phi$, 
all possible combinations of BDD results
must be taken into consideration when solving this problem\footnote{Each 
possible combination of BDD results is a unique binary vector $\bm \eta=\{\eta_{\bm\phi}\}_{\forall \bm \phi}$,
which results in a different objective function.}.
For example, in a distribution network with $E+U$ sensors, including PMUs and EV counters,
there are $2^{E+U}$ possible communication results, each resulting in a specific $\bm z^{AR}$.
As a result, there are $2^{2^{E+U}}$ possible BDD results (i.e., $\bm\eta$ vectors)
because each $\bm z^{AR}$ either passes BDD or it does not.

Observe that the objective function $\Psi(\bm\alpha)$ of~\eqref{original-problem} 
is the weighted sum of several $\psi_{\bm\phi}(\bm \alpha)$ terms, 
each being similar to the objective function of \eqref{subproblem}, but for a specific $\bm\phi$.
Thus, if the values of $\eta_{\bm\phi}(\bm\alpha)$ elements are fixed, 
the objective function $\Psi(\bm\alpha)$ becomes a linear function of $\bm\alpha$
and~\eqref{detect} can be converted to an inequality constraint for each $\phi$
in the form of $\|\bm r^{A}_{\bm\phi}\|_2\leq \epsilon$.
% $g^R_{\bm\phi}(\alpha)=||\bm r^{A}_{\bm\phi}||^2_2-\epsilon^2\leq 0$, 
% which is an ellipsoid in a high-dimensional space.
This implies that the feasible set is the intersection of 
$2^{E+U}$ ellipsoids which are obtained by squaring the constraints.
Hence, for each vector $\bm \eta$, the optimization problem is a convex 
quadratically constrained linear program (QCLP).
It can be further shown that this QCLP is a special case of a second order cone program (SOCP)
and can therefore be solved efficiently by interior point methods~\cite{cvopt}.

Algorithm~1 describes how~\eqref{original-problem} is solved 
by iterating over the set of $\bm \eta$ vectors
and solving the resulting QCLP in each case.
The maximum adverse impact $\Psi(\bm\alpha)$, attained at the solutions of these problems,
determines the solution of the original stochastic optimization problem.
Note that in Line~4, 
we check whether the feasible set is empty and discard $\eta$ if this is the case. 
This is because an empty feasible set indicates that 
no attack vector can lead to this specific combination of BDD results 
under an arbitrary communication result.

\begin{algorithm}[t!]
	\caption{FDIA Vector Construction}
	\LinesNumbered 
	\KwIn{$\mathbf H$, $\mathbf W$, $\bm z$, $\bm z^P$, $ \Omega$, $\mathcal V$, $\epsilon$, $ \pi(G)$, $\pi(B)$}
	\KwOut{$\bm \alpha^*$}
	%\For{each vector $\bm \eta$}{
	%\For{each vector $\bm \eta\in\{[\overbrace{\cdots,\eta_i,\cdots}^{2^{E+U}}]|\eta_i\in\{0,1\}\}$}{
	$\alpha_{\bm \eta}^*\gets 0$ \tcp*{Initialization}
	\For{each vector $\bm \eta$}{
	Solve the respective convex QCLP\;
	\If{Optimal point (denoted $\bm\alpha_{\bm\eta}'$) exists \textbf{and} $\Psi(\bm \alpha_{\bm\eta}') > \Psi(\bm \alpha_{\bm\eta}^*)$}
				{$\bm \alpha_{\bm\eta}^* \gets \bm \alpha_{\bm\eta}'$\;}
	}
% $\bm \alpha^*=\max\{\bm \alpha^*,\max\{\bm\alpha_{\bm\eta}^*\}\}$\;
%Return $\bm \alpha^*$
Return $\bm\alpha_{\bm\eta}^*$
\label{alg}\end{algorithm}

\section{Case Studies}
\begin{figure}[!t]
	\centering
	\includegraphics[width=\linewidth]{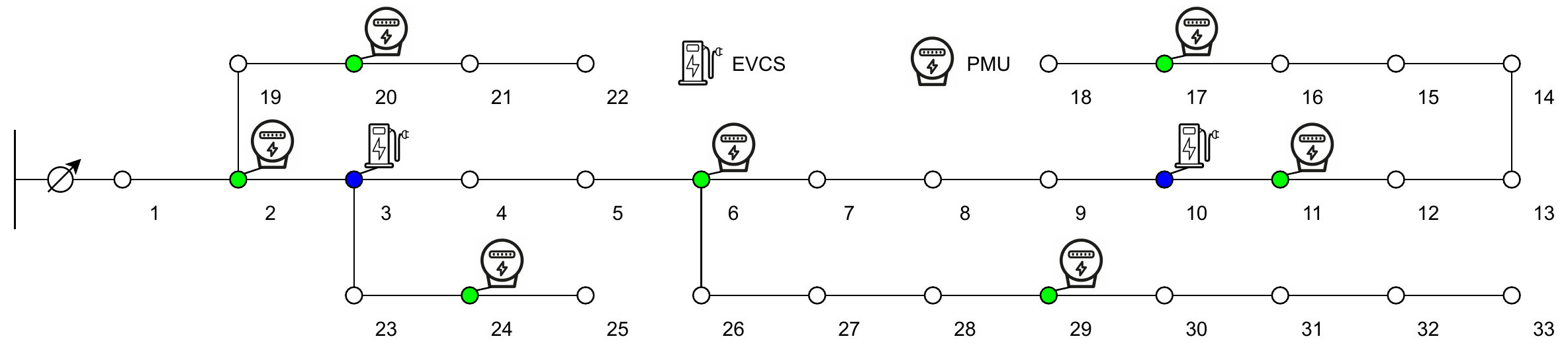}
	\caption{IEEE 33-bus test feeder topology}
	\label{fig:ieee33bus}
\end{figure}

\begin{figure}[!t]
	\centering
	\includegraphics[width=\linewidth]{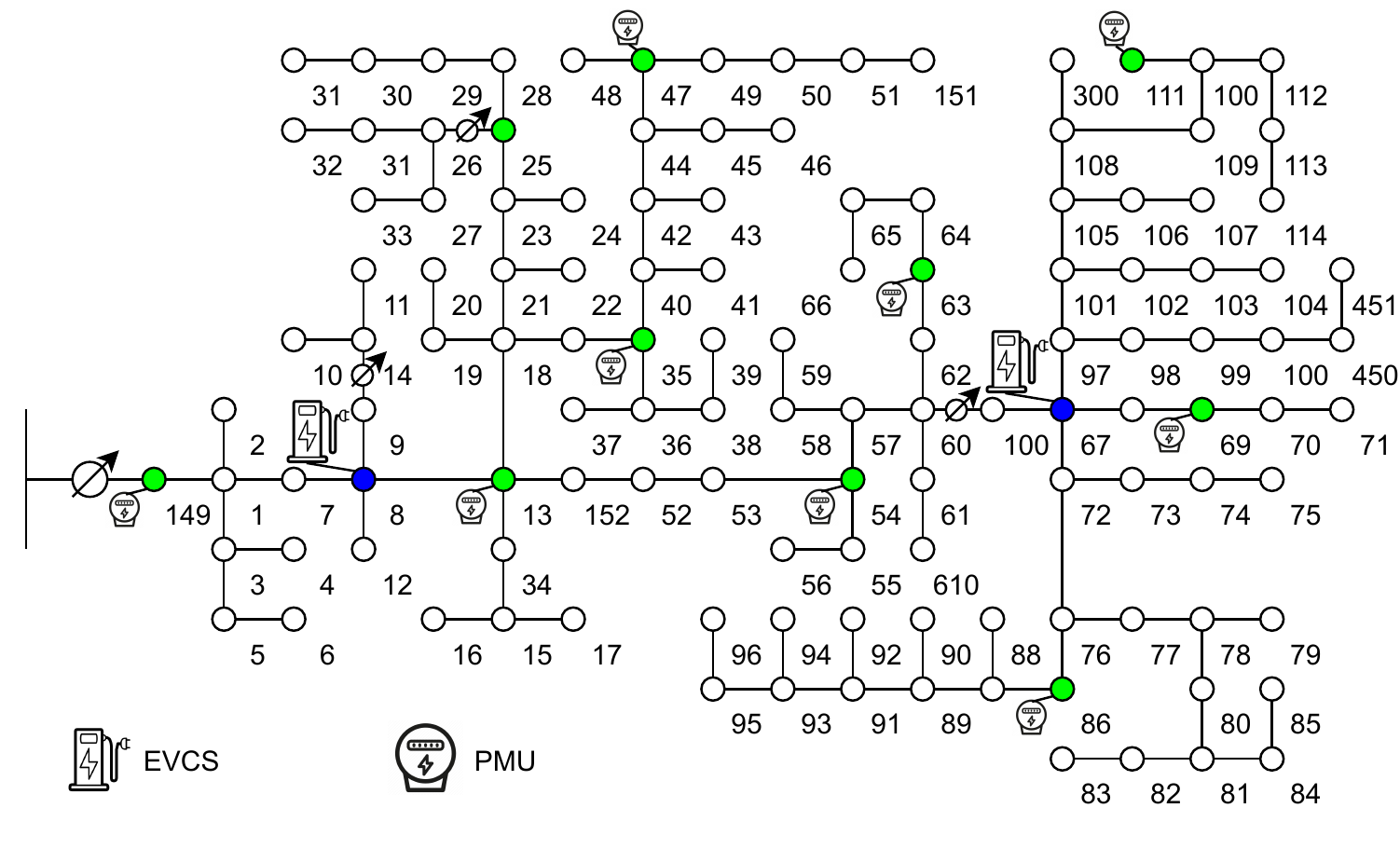}
	\caption{IEEE 123-bus test feeder topology}
	\label{fig:ieee123bus}
\end{figure}

To investigate the impact of stochastic communication on the FDIA performance 
and the vulnerability of the distribution system VR process to this attack,
we evaluate the proposed FDIA on the IEEE 33-bus test feeder~\cite{bus33} 
and a simplified version of the IEEE 123-bus test feeder~\cite{bus123}. 
Fig.~\ref{fig:ieee33bus} and \ref{fig:ieee123bus} depict 
where charging stations and distribution-level PMUs are installed in these systems.
To guarantee accurate state estimation, 
the minimum number of PMUs should be between 1/5 and 1/3 of the total number of buses~\cite{pmunum}.
Thus, we consider a total of 7 PMUs in IEEE 33-bus test feeder 
and 10 PMUs in the simplified IEEE 123-bus test feeder. 
The pseudo measurement of PMUs can be obtained from the historical data if needed. 
The pseudo measurement of EVCS counters are obtained 
from the EVCS queuing model presented in Section~\ref{sec:queueing}.  
Following~\cite{std}, we set the error of magnitude and phase angle measurements 
to be an additive white Gaussian noise, $N(0,0.01)$ and $N(0,0.005)$, respectively. 
The base loads and PMU pseudo measurements are generated according to~\cite{openei}, 
and the pseudo measurement error is assumed to be 
an additive white Gaussian noise $N(0,0.09)$~\cite{pseudostd}. 


To simulate the stochastic communication process, %and investigate the impact on VR process and FDIA strategy, 
we use an open-source co-simulation platform developed in~\cite{cosimu}. %based on Mosaik.
In this platform, OpenDSS is utilized for power flow analysis (PFA) 
and a network simulator component is used to capture the communication bit error rate. 
The bit error rate is set to 0.01, the communication involves small frames with a payload of 32 bits, 
and framing overhead is considered negligible. 
The communication network topology is assumed to be a star, 
with the DSO at the center of the star.

There are 2 charging stations in each distribution system. 
%each station is equipped with an EV counter. 
It is assumed that each EV counter sends the EV count
in the respective EVCS to the DSO via the same communication network as PMUs.
% and the measurement error of the EV counter is zero. 
% To simulate the daily operation of an EVCS, 
We use real data to generate the initial SOC, charging demand, and parking time of EVs that visit an EVCS. 
Specifically, the charging demand is approximated based on the product of trip distance, 
in the NHTS dataset~\cite{nhts}, and the average energy consumption per mile 
together with the battery size and corresponding maximum driving mileage obtained from~\cite{evd}. 
The charging power and parking time are obtained from the EVnetNl dataset~\cite{evnetnl}. 
A log-normal distribution with expected value of 0.6848 and standard deviation of 0.9353 
is fitted to the empirical distribution of parking times.
The charging power is divided into 23 discrete levels from 1 to 23 kW, 
and the empirical probability mass function is obtained from the dataset.
The initial SOC of EVs is generated given the battery size and daily trips. 
In each EVCS, 50 parking stalls are assumed to be equipped with chargers 
that support charge powers ranging from 1 to 23~kW.

\begin{table}[t!]
	\caption{Mean Absolute Percentage Error of $\Delta v_{\bm\phi}$}
	\label{tabel:esterr}
	\begin{tabular}{l|cc}
		\hline
		Circuit                  & Nonlinear PFA & Linear PFA \\ \hline
		IEEE 33-bus test feeder  & 0.714\%       & 1.654\%    \\
		IEEE 123-bus test feeder & 1.154\%       & 1.916\%    \\ \hline
	\end{tabular}
\end{table}
We study how the proposed FDIA affects the VR process and 
compare its performance assuming stochastic communication (SC) 
with the traditional FDIA that builds on an idealized communication (IC) model~\cite{stateest4}. 
Three cases will be mainly discussed: original case without FDIA, FDIA with IC, and FDIA with SC.
Before discussing the FDIA performance, 
we look at the error introduced in the original case in the DSSE process, 
which is repeated once every 10~minutes (144 runs in a day), 
due to linearizing power flow equations (Section~\ref{sec:linpf}).
Both the nonlinear PFA and linearized PFA approaches are tested through co-simulation
and the mean absolute percentage error (MAPE) of the VR capacity estimation 
in each case is reported in Table~\ref{tabel:esterr}. 
It can be readily seen that the approximation done in the linearized PFA
does not introduce significant additional error compared to the nonlinear analysis. 
%Note that the error due to the approximation can be larger in the IEEE 123-bus test feeder than in the IEEE 33-bus test feeder, which implies that it is necessary to check the estimation error  before using the linearized power flow equations in a larger distribution system.

\begin{figure}[t]
	\centering
	\includegraphics[width=\linewidth]{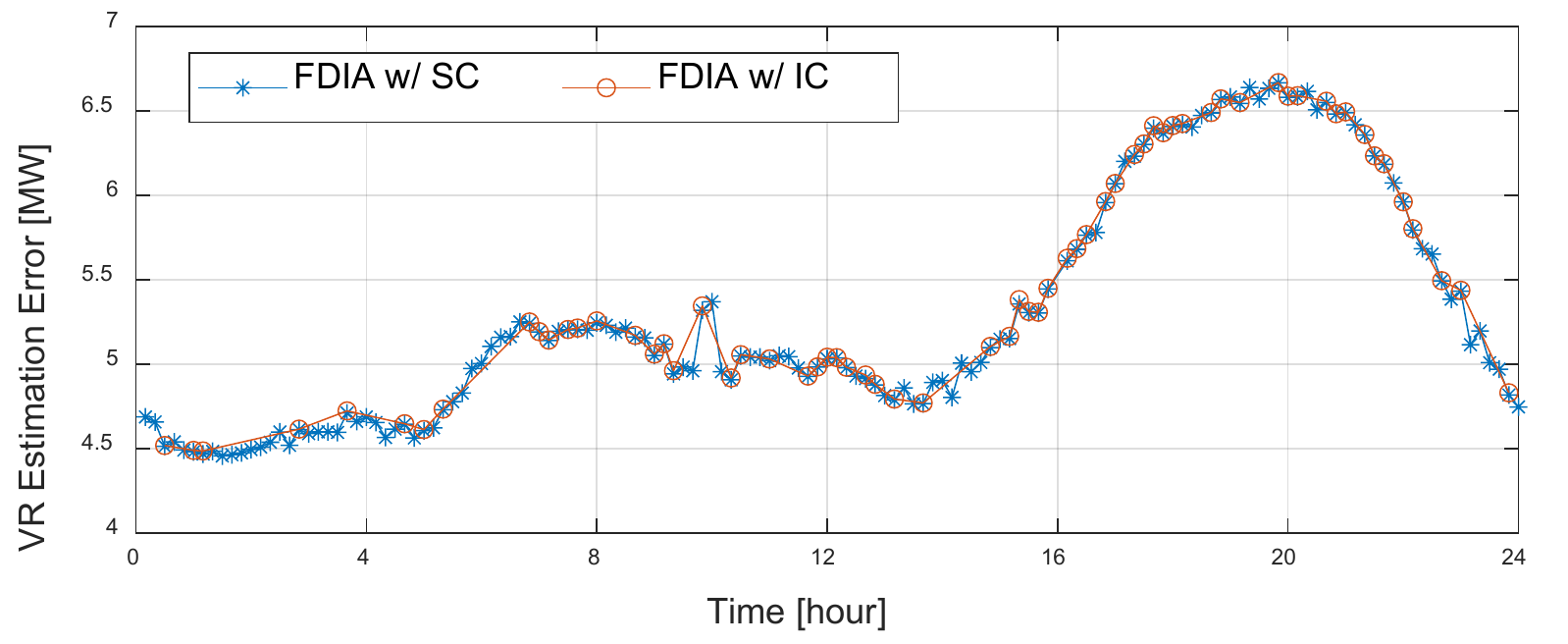}
	\caption{Comparison of VR capacity estimation error in the IEEE 33-bus system.
	Note that a dot/marker is drawn only when the attack vector bypasses BDD.}
	\label{fig:bus33estimationerror}
\end{figure}

We now evaluate the performance of the proposed FDIA with SC 
in a case study that involves the IEEE 33-bus test feeder, 
where an FDIA vector is computed in every time slot (10~min) 
and the total duration of our simulation is one day. 
%Fig.~\ref{fig:bus33estimationerror}~and~\ref{fig:bus33cdp} show the results. 
Fig.~\ref{fig:bus33estimationerror} shows the VR estimation error caused by the FDIA vectors bypassing the BDD. 
We can see that FDIAs can cause a noticeable VR capacity estimation error, i.e., $\psi(\bm \alpha)$. 
Compared to FDIA with IC, the proposed FDIA (with SC) can effectively mislead the DSO,
resulting in similar VR capacity estimation errors with higher BDD pass rate\footnote{There are many more blue markers
than red markers which indicate that more attack vectors did not 
pass BDD in the case of FDIA with IC.}.
In particular, the FDIA with SC increases the MAPE of the VR capacity estimation to 427\% 
and the FDIA with IC increases it to 433\%, 
which is slightly higher but results in a much higher detection rate.
% Although in a small number of time slots
% FDIA with IC may cause a larger VR capacity estimation error, 
% in general the FDIA with SC has a greater probability of causing a large estimation error.

\begin{figure}[t]
	\centering
	\includegraphics[width=\linewidth]{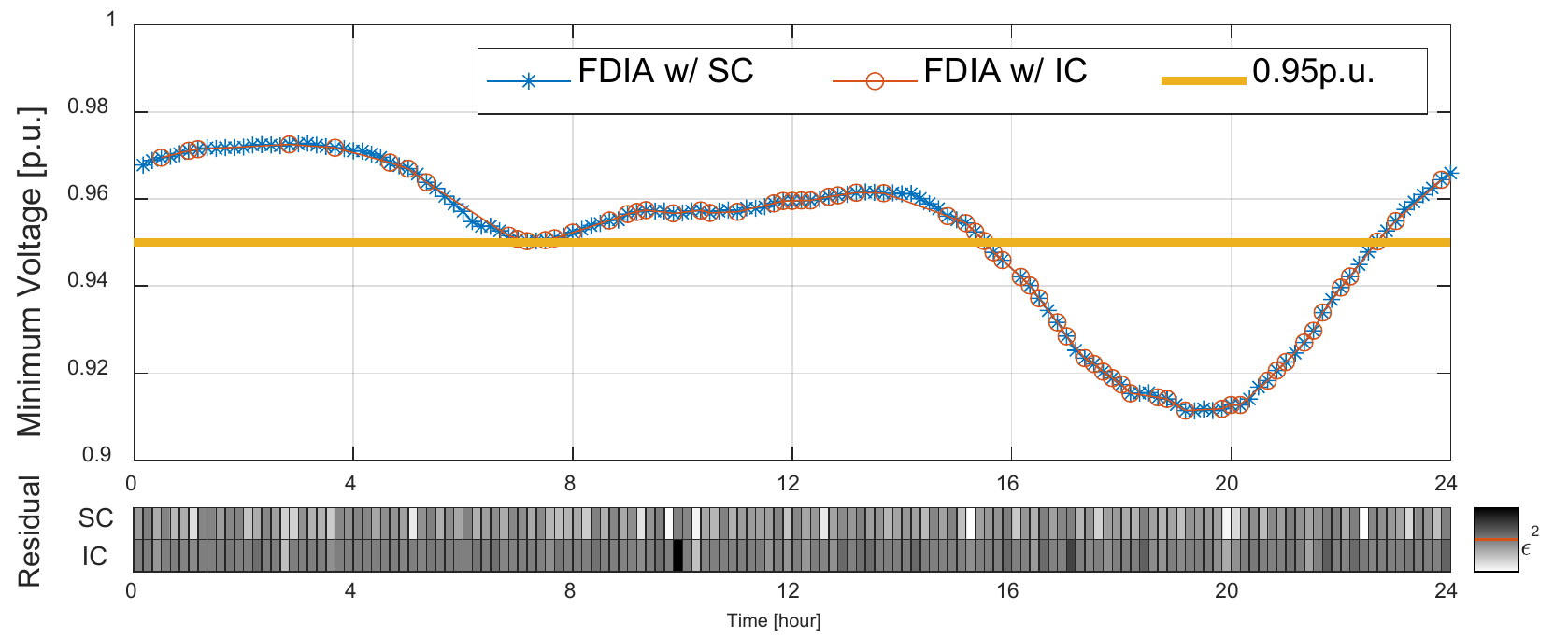}
	\caption{Comparison of minimum voltage magnitude and BDD pass rate in the IEEE 33-bus system.
	The heatmaps below this figure show the value of $\|\bm r^{A}_{\phi}\|_2^2$ and the horizontal (red) line 
	in the colorbar marks the color intensity of $\epsilon^2$. 
	Thus, any point that is painted with a lighter gray corresponds to an attack that has bypassed BDD.}
	\label{fig:bus33minvol}
\end{figure}

\begin{figure}[t!]
	\centering
	\includegraphics[width=\linewidth]{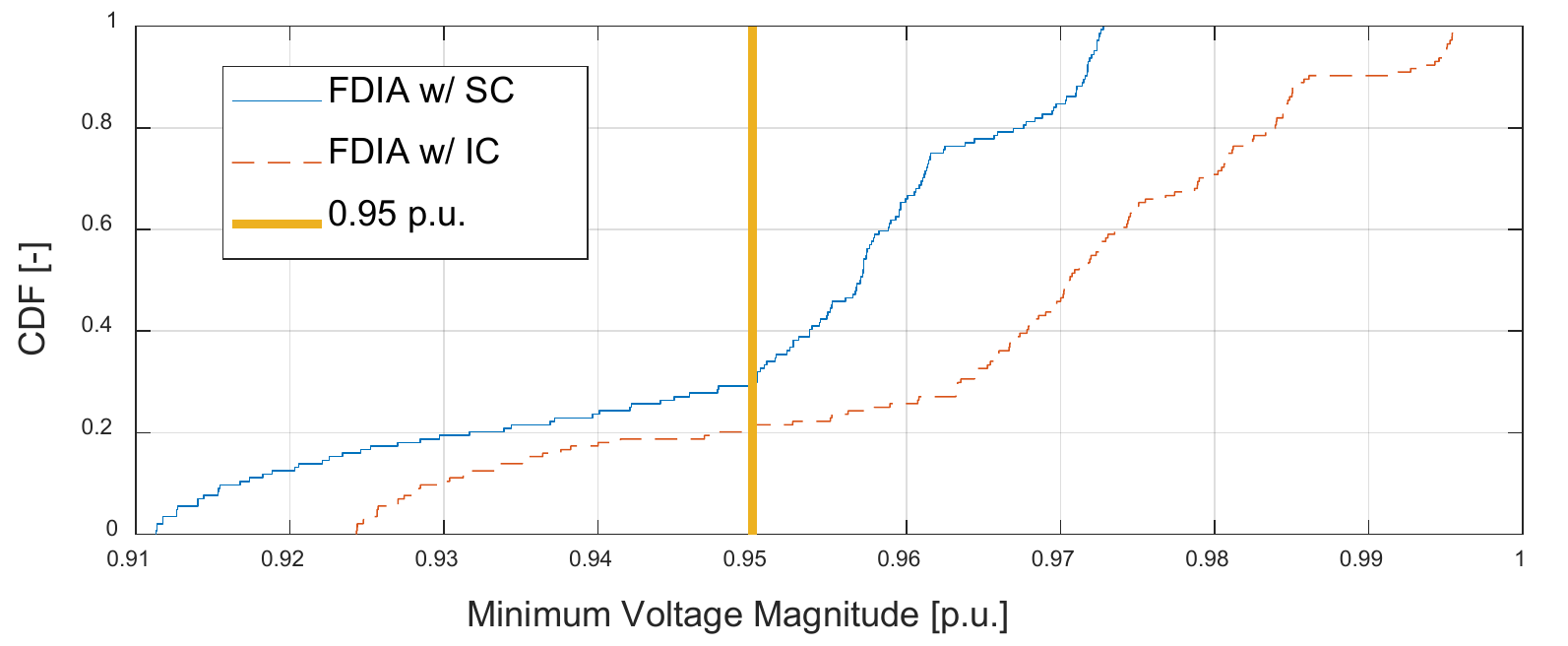}
	\caption{Cumulative distribution function of the minimum voltage magnitude 
	in the IEEE 33-bus system.}
	\label{fig:bus33cdp}
\end{figure}

In addition to the increased VR capacity estimation error, 
the negative impact of FDIAs on the voltage profiles is also more pronounced. 
As shown in Fig.~\ref{fig:bus33minvol}, both FDIAs computed by the attacker 
can result in under-voltage incidents (voltage dropping below $0.95pu$). 
This is while there is no under-voltage incidents without FDIA 
because the DSO can accurately estimate the EVCS VR capacity 
and the insufficient capacity is met by other VR resources. 
But, the DSO issues an erroneous VR request under FDIA. %for the EVCSs. 
Since the EVCS VR capacity is overestimated and 
the actual VR capacity is not enough to satisfy the VR request, 
serious damages can be inflicted on the power system. %due to the FDIA. 
Moreover, considering the randomness of the communication process, 
the proposed FDIA with SC causes 41 under-voltage incidents during one day, 
which is approximately twice the number of incidents seen in the case of FDIA with IC. 
%We can explain this by inspecting the BDD pass rate profile. 
This can be attributed to the higher number of modified measurements 
that manage to go past BDD as can be seen in the bottom subplot of Fig.~\ref{fig:bus33minvol} (the two heatmaps).
Recall that not all injected false data can be received by the DSO in time 
for VR optimization due to the stochastic packet drops that occur in the communication network. 
Thus, the attack vector that bypasses BDD mechanisms in theory might be actually detected. 
By considering the packet drop probability, 
the proposed FDIA can increase the BDD pass rate from around 45\% to 99.3\% 
(i.e., one under-voltage incident in 144 time slots);
this greatly increases the stealthiness of FDIA. 
From Fig.~\ref{fig:bus33cdp} we can also conclude that 
the proposed FDIA is more likely to create under-voltage incidents than FDIA with IC. 


\begin{figure}[t]
	\centering
	\includegraphics[width=\linewidth]{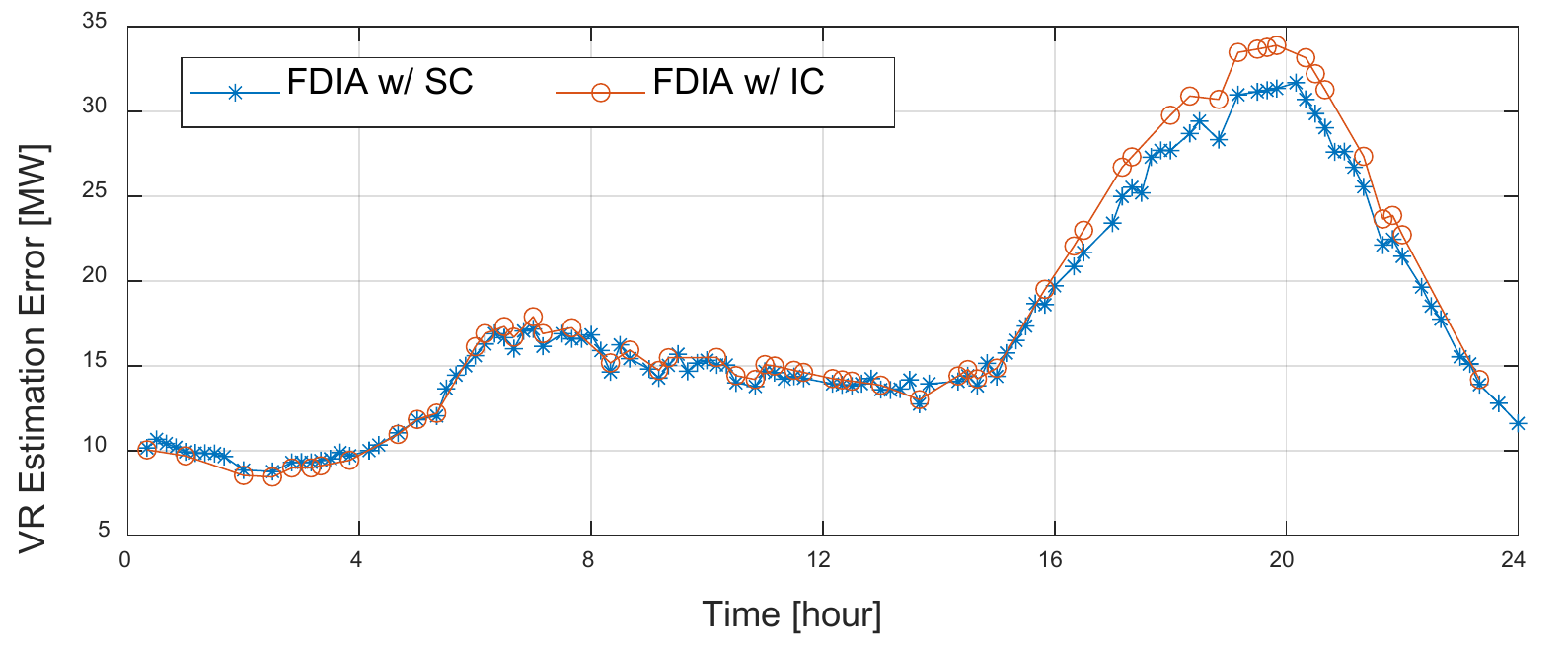}
	\caption{Comparison of VR capacity estimation error in the IEEE 123-bus system.
	Note that a dot/marker is drawn only when the attack vector bypasses BDD.}
	\label{fig:bus123estimationerror}
\end{figure}

\begin{figure}[t]
	\centering
	\includegraphics[width=\linewidth]{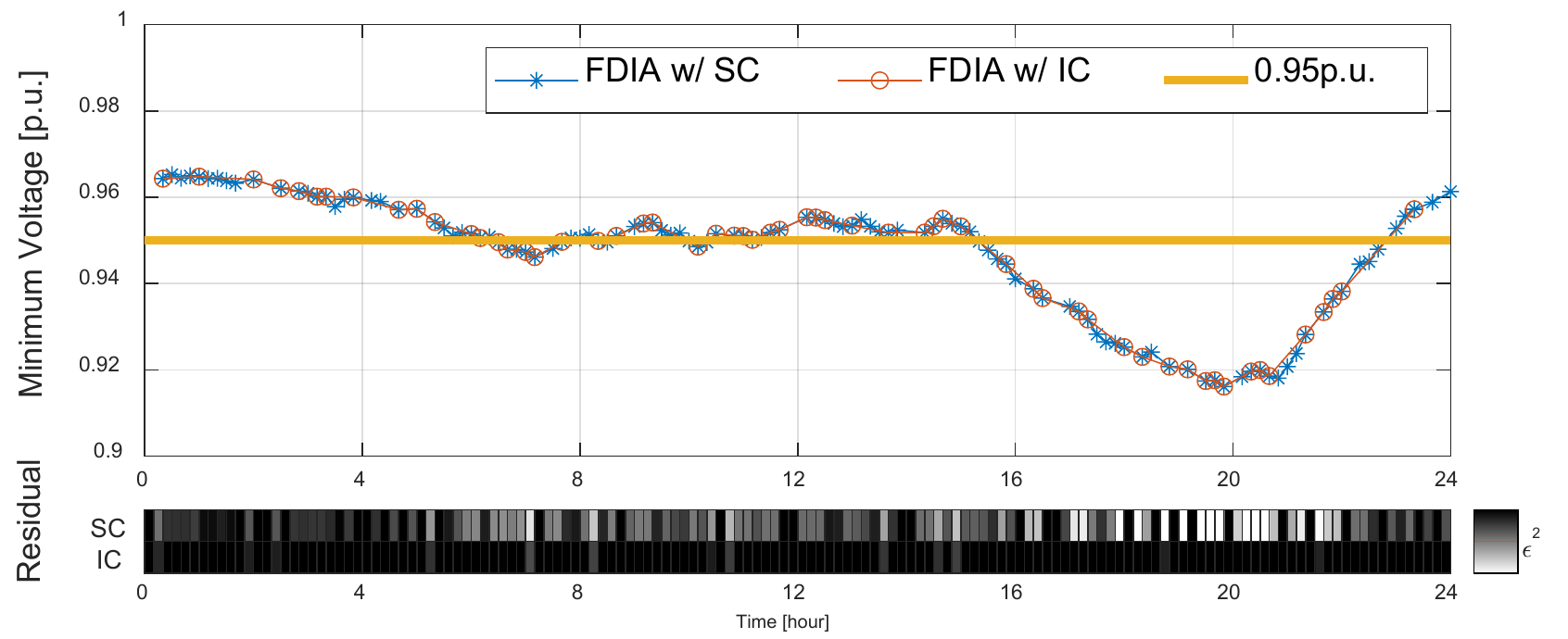}
	\caption{Comparison of minimum voltage magnitude and BDD pass rate in the IEEE 123-bus system.
	The heatmaps below this figure show the value of $\|\bm r^{A}_{\phi}\|_2^2$ and the horizontal (red) line in the colorbar 
	marks the color intensity of $\epsilon^2$. 
	Thus, any point that is painted with a lighter gray corresponds to an attack that has bypassed BDD.}
	\label{fig:bus123minvol}
\end{figure}

\begin{figure}[t!]
	\centering
	\includegraphics[width=\linewidth]{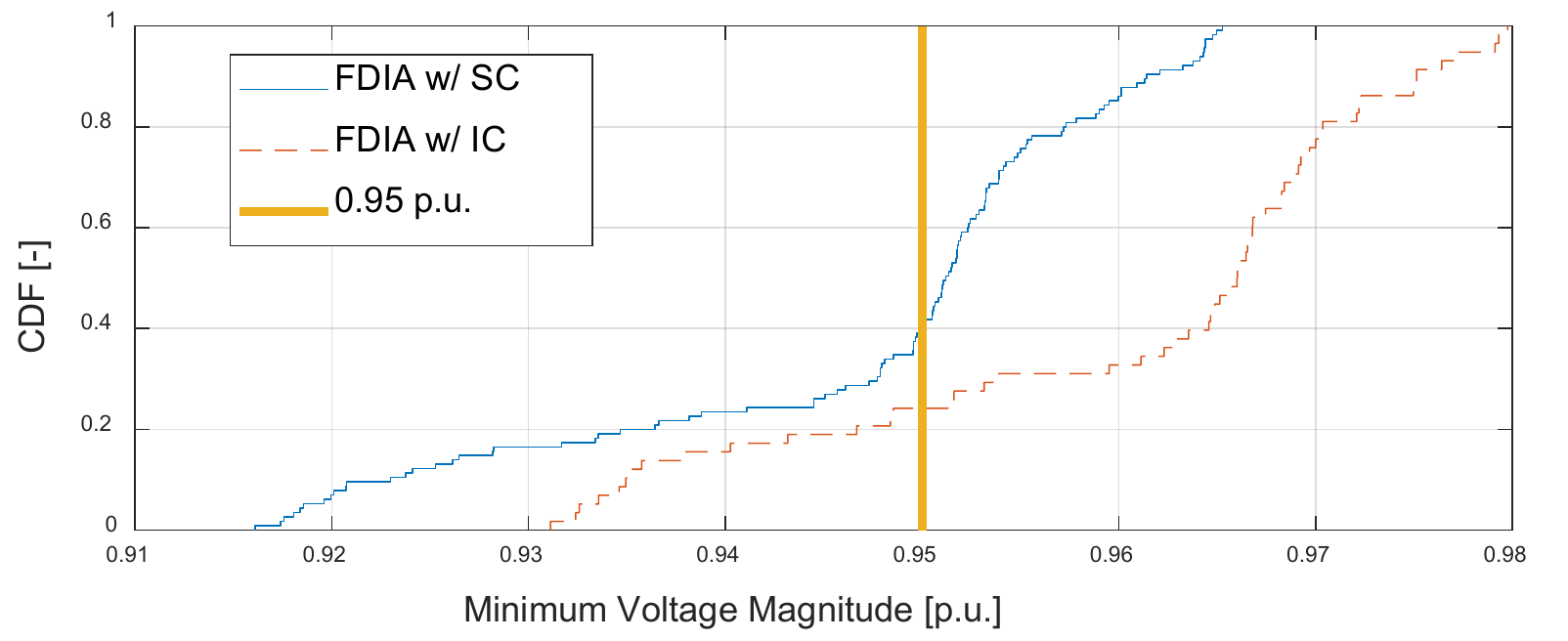}
	\caption{Cumulative distribution function of the minimum voltage magnitude 
	in the IEEE 123-bus system.}
	\label{fig:bus123cdp}
\end{figure}

Next, we turn our attention to the case study that involves the IEEE 123-bus test feeder. 
The goal is to show that in a large distribution system, 
the performance of the proposed FDIA does not fall apart. 
As it can be seen in Fig.~\ref{fig:bus123estimationerror}, 
the MAPE of VR capacity estimation caused by FDIA with SC is 674\%, 
which is on par with the 691\% relative error caused by FDIA with IC. 
%The high probability of causing a large estimation error through the proposed FDIA  is also shown in Fig.~\ref{fig:bus123estcumuerr}. 
% For the same number of charging stations and the same number of charging points per station, 
% the VR capacity estimation error in the IEEE 123-bus test feeder 
% is almost the same as that in the IEEE 33-bus test feeder. 
% Yet, since the VR requirements are higher in the 123-bus test feeder, 
% more under-voltage incidents are caused by the FDIA. 
We witness that FDIA with IC introduces a higher error than FDIA with SC (especially from 16:00 to 22:00), 
but this comes at the cost of being detected.
As shown in Fig.~\ref{fig:bus123minvol}, 
the FDIA with IC causes 26 under-voltage incidents, while the proposed FDIA causes 47 incidents. 
By inspecting the BDD pass ratio and the corresponding under-voltage incidents, 
it is evident that the attacker can also achieve a better performance 
when it considers the stochasticity of data communication. 
The greater potential of the proposed FDIA to cause more detrimental under-voltage incidents 
in larger distribution systems is depicted in Fig.~\ref{fig:bus123cdp}. % relatively higher, which is shown in Fig.~\ref{fig:bus123cdp}.


\section{Conclusion and Future Work}
This paper proposes a novel FDIA against EVCS-mediated voltage regulation
which has been the subject of several studies recently.
This attack considers potential delays and packet losses in the communication network
and the stochastic mobility pattern and charging demand of an EV fleet
to maximize its expected adverse impact on the distribution system over time.
We carry out simulation on two standard test feeders 
using a co-simulation platform to showcase the 
greater potential of this FDIA to inflict damage
compared to the state-of-the-art FDIA attack that relies on an idealized communication model.
Our result highlights the vulnerability of the existing BDD mechanism
that protects the DSSE process.

In future work, we intend to extend the BDD mechanism to address this vulnerability.
A possible approach to detect such an attack is for 
the DSO to consider various communication results in a network simulator 
to decide if the received measurement can be trusted.
This can help to improve the detect rate 
and complicate the attack vector construction process, 
thereby preventing the attacker from solving it in a timely fashion.
Furthermore, we plan to develop a probabilistic model 
for the participation of EVs as economic resources in the VR process. 
This model will capture the battery degradation cost and 
incorporate the monetary incentive that will be provided to individual EVs.

% Besides, observer that such a FDIA construction approach highly depends on the estimation of communication results $\eta_{\phi}$. Hence, to detect the proposed FDIA, one possible solution is to test the received measurements with various communication results in a network simulation environment, where the DSO can assume multiple communication results $\eta_{\phi}$ for the DSSE and BDD process. Since these cases have not been taken into account according to \eqref{detect}, the detection rate for the proposed FDIA can be improved. Moreover, the BDD mechanism in \eqref{residual} and \eqref{detect} can be modified with more complex limitations. In this way, the optimization problem of FDIA vector construction will be too complex to be solved, which could also protect the VR process from the proposed FDIA effectively.

\end{document}